\documentclass{aastex63}

\usepackage[utf8]{inputenc}

\shorttitle{Citizen ASAS-SN}
\shortauthors{Christy et al.}
\graphicspath{{./}{figures/}}

\begin{document}

\title{Citizen ASAS-SN: Citizen Science with The All-Sky Automated Survey for SuperNovae (ASAS-SN)}

\author[0000-0003-0528-202X]{C. T. Christy}
\affiliation{Department of Astronomy, The Ohio State University, 140 West 18th Avenue, Columbus, OH 43210, USA}

\author[0000-0002-6244-477X]{T. Jayasinghe} 
\affiliation{Department of Astronomy, The Ohio State University, 140 West 18th Avenue, Columbus, OH 43210, USA}

\author{K. Z.  Stanek}
\affiliation{Department of Astronomy, The Ohio State University, 140 West 18th Avenue, Columbus, OH 43210, USA}

\author{C. S. Kochanek}
\affiliation{Department of Astronomy, The Ohio State University, 140 West 18th Avenue, Columbus, OH 43210, USA}

\author{Z. Way}
\affiliation{Department of Astronomy, The Ohio State University, 140 West 18th Avenue, Columbus, OH 43210, USA}

\author{J. L. Prieto}
\affiliation{Núcleo de Astronomía,
Universidad Diego Portales, Av. Ejército 441, Santiago,
Chile}

\author{B. J. Shappee}
\affiliation{Institute for Astronomy, University of Hawai’i, 2680 Woodlawn Drive, Honolulu, HI 96822,USA}

\author[0000-0001-9206-3460]{T. W.-S. Holoien}
\altaffiliation{NHFP Einstein Fellow}
\affiliation{The Observatories of the Carnegie Institution for Science, 813 Santa Barbara Street, Pasadena, CA 91101, USA}

\author{T. A. Thompson}
\affiliation{Department of Astronomy, The Ohio State University, 140 West 18th Avenue, Columbus, OH 43210, USA}

\begin{abstract}

We present ``Citizen ASAS-SN", a citizen science project hosted on the Zooniverse platform which utilizes data from the All-Sky Automated Survey for SuperNovae (ASAS-SN). Volunteers are presented with ASAS-SN $g$-band light curves of variable star candidates. The classification workflow allows volunteers to classify these sources into major variable groups, while also allowing for the identification of unique variable stars for additional follow-up. 

\end{abstract}

\section{Introduction}
ASAS-SN is a wide-field photometric survey that monitors the entire night sky using 20 telescopes located in both hemispheres \citep{2014ApJ...788...48S,2017PASP..129j4502K}. The field of view of an ASAS-SN camera is 4.5 deg$^2$, the pixel scale is 8\farcs0 and the FWHM is ${\sim}2$ pixels. ASAS-SN uses image subtraction \citep{1998ApJ...503..325A} for the detection of transients and to generate light curves. Recently, we have been using ASAS-SN data to study bright variable stars (see, for e.g., \citealt{Jayasinghe_2021}). In the \textit{V}-band catalog, ${\sim} 60$ million stars were classified using machine learning, resulting in a catalog of ${\sim}426,000$ variables, of which ${\sim} 220,000$ were new discoveries. Since 2018, ASAS-SN has shifted to the \textit{g-}band with a depth of $g\lesssim18.5$ mag, allowing for up to 100 million stars to be characterized. The new \textit{g-}band data has an improved cadence ($\lesssim 24$ hours, vs. ${\sim}2-3$ days $V$-band) and reduced diurnal aliasing due to the longitudinal spread of the ASAS-SN units. 

While machine learning methods are efficient at analyzing known behaviors, the scarcity of rare phenomena makes it difficult for them to account for new or extreme cases \citep{1410299}. One alternative approach is citizen science, which enables volunteers to participate in the scientific process. In recent years, the Zooniverse\footnote{Zooniverse:https://www.zooniverse.org/} has hosted many successful citizen science projects. Citizen scientists working with \emph{Planet Hunters} \citep{2016MNRAS.457.3988B} found KIC8462852 (aka Boyajian's star), which continues to be studied \citep{2018JAVSO..46...14S,2020arXiv200506569H}. The \emph{Galaxy Zoo} project used citizen science to morphologically classify galaxies and has produced over 60 publications, millions of classifications, and multiple spin-off projects \citep{2020IAUS..353..205M}. One of the most unusual objects was an emission nebula near IC 2497 known as ``Hanny’s Voorwerp" \citep{2009A&A...500L..33J}.

\section{Citizen ASAS-SN} \label{sec:floats}
ASAS-SN's citizen science project focuses on the classification of \textit{g-}band light curves to identify both classical and anomalous variables. The identification and study of unusual variable stars often lead to new astrophysical insights. Examples from ASAS-SN include the most extreme heartbeat star found to date \citep{Jayasinghe_2019}, and a Tabby's star analogue (dubbed Zachy's star, \citealt{2019ATel13346....1W}).

\begin{figure}[t]
\begin{center}
       \includegraphics[width=1.0\textwidth]{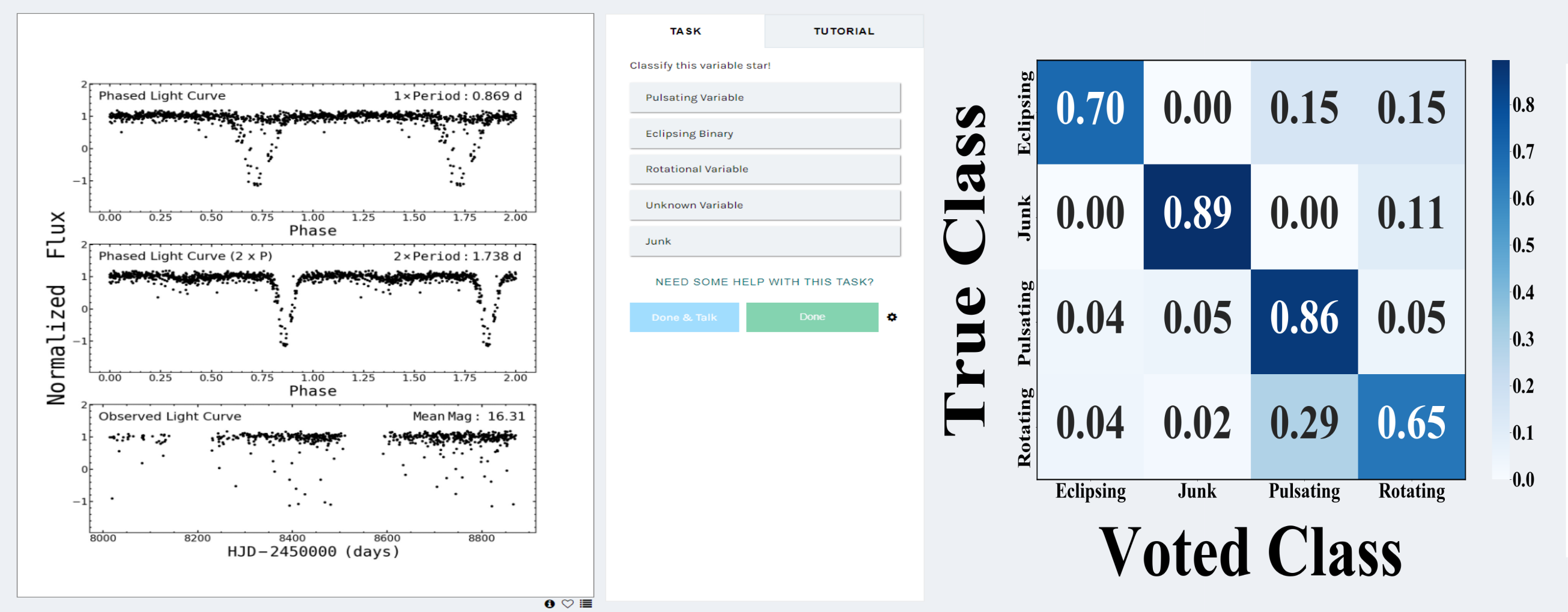}
 	   \caption{(Left) A snapshot of the Citizen ASAS-SN classification workflow. For this particular candidate, the correct classification is \emph{Eclipsing Binary}. (Right) The confusion matrix from our internal review.}
    \label{fig:fig1} 	   
\end{center}
\end{figure}

We used the \verb"refcat2" catalog \citep{2018ApJ...867..105T} as our catalog of targets. The $g$-band light curves for candidate variables were extracted as described in \citet{Jayasinghe_2021}. We corrected the zero point offsets between the different cameras as described in \citet{Jayasinghe_2021} and calculated periodograms using the Generalized Lomb-Scargle (GLS, \citealt{2009A&A...496..577Z,1982ApJ...263..835S}) periodogram. We display the original light curve along with light curves phased by both the best GLS period and twice that period (Figure \ref{fig:fig1}).

As a first foray into citizen science, our goal was to address several simple but common classification problems. In particular, ambiguities arise for certain variable types such as RRC RR Lyrae v.s. EW eclipsing binaries. For eclipsing binaries, the best period returned by the GLS periodogram is often 1/2 the orbital period, which is why we show two phasings. Using the correct orbital period, eclipsing binaries show a distinct separation of the primary and secondary eclipses allowing for their accurate classification. The observed light curve is useful for identifying long-period variables, and evolving variables like rotating spotted stars.

We designed our workflow to be easy to navigate and accessible to a wide array of volunteers. Because we expect no prior knowledge of variable star classification, we first present users with a tutorial that details the classification process and briefly summarizes the science. Volunteers also have access to a field guide describing common variable stars and their light curves. Our workflow tasks users to determine the correct basic classification, selecting between three broad classes (Pulsating Variables, Eclipsing Binaries, Rotational Variables), choosing the option ``Unknown Variable" for ambiguous cases, or flagging it as ``junk". As users begin, we present them with a variety of ``gold-standard" (GS) candidates classified by ASAS-SN. These GS variables provide the user with feedback on their classifications to train them in the process. As users make more classifications, GS variables become less frequent and the user begins to classify new light curves.

In preparation for launch, we submitted Citizen ASAS-SN to the Zooniverse team for their internal review. During this phase, 58 volunteers from Zooniverse tested the workflows and provided feedback on their experience. Many users suggested simplifying the tutorial and including more examples in the field guide, leading to an update. During this stage, users classified a set of 200 new candidates and $\sim$ 400 GS variables. Figure \ref{fig:fig1} shows the resulting confusion matrix; the ``voted" class represents the final majority vote that a candidate received while the ``true class" indicates the correct classification. When we analyzed the classifications, we found that users could reliably distinguish between the three main types. Individuals correctly identified 86\% of pulsating variables, along with 70\% of eclipsing binaries, 65\% of rotational variables, and 89\% of variables labeled as ``junk" by our team. For the non-GS variables, the classification accuracy for pulsating variables and eclipsing binaries remained high; however, rotational variable classifications were systematically less accurate. Additionally, we found that $\sim 20 \%$ of our 200 test variables were classified as junk. We also removed the second task for eclipsing binaries which asked users to determine if period doubling was occurring. Users were quick to label binaries as having the original GLS period, when in fact most of the binaries in our sample had twice this period. We suspect that this question is better addressed in a workflow dedicated solely to eclipsing binaries.

As we progress with the project, we plan to introduce other workflows, including classifications of irregular variables and higher-order classifications of specific variable types. For the variables that were studied using the \textit{V-}band data, we will compare the classifications made by Zooniverse volunteers to those obtained through our machine learning classifiers, and update the ASAS-SN variable stars database\footnote{ASAS-SN variable stars database:https://asas-sn.osu.edu/variables} with improved classifications as needed.

\vspace{.125in}
ASAS-SN is supported at OSU by the Gordon and Betty Moore
Foundation through grant GBMF5490 to the Ohio State University and NSF grant AST-1908570.

\bibliographystyle{aasjournal}
\bibliography{mybib}

\end{document}